\documentclass[9pt,twocolumn,twoside]{optica-article}

\journal{opticajournal} % for journals or Optica Open

\articletype{Research Article}

\usepackage{lineno}
\usepackage{comment}
\usepackage{bm}
\usepackage{float}
\usepackage{dcolumn}
\usepackage{soul}
\usepackage{cancel}
\usepackage[normalem]{ulem} % Load ulem package for strikethrough
\usepackage[theorems,skins]{tcolorbox}
\newtcolorbox{mymathbox}[1][]{colback=white, sharp corners, #1}
\usepackage{graphicx}
\usepackage[margin=0cm]{caption}

\begin{document}

\title{Enhanced fidelity in nonlinear structured light by virtual light-based apertures}

\author{Sachleen Singh\authormark{1}, Isaac Nape\authormark{1}, Andrew Forbes\authormark{1,*}} 

\address{\authormark{1} School of Physics, University of Witwatersrand, Johannesburg }
\email{\authormark{*}andrew.forbes@wits.ac.za} 

\begin{abstract*} 
Tailoring the degrees of freedom (DoF) of light for a desired purpose, so-called structured light, has delivered numerous advances over the past decade, ranging from communications and quantum cryptography to optical trapping, and microscopy. The shaping toolkit has traditionally been linear in nature, only recently extended to the nonlinear regime, where input beams overlap in a nonlinear crystal to generate a structured output beam.  Here we show how to enhance the fidelity of the structured output by aligning light with light. Using orbital angular momentum modes and difference frequency generation as an example, we demonstrate precise control of the spatial overlap in both the transverse and longitudinal directions using the structure of one mode as a virtual structured (in amplitude and phase) light-based aperture for the other. Our technique can easily be translated to other structured light fields as well as alternative nonlinear processes such as second harmonic generation and sum frequency generation, enabling advancements in communication, imaging, and spectroscopy.

\end{abstract*}

%%%%%%%%%%%%%%%%%%%%%%%%%%  body %%%%%%%%%%%%%%%%%%%%%%%%%%
\section{Introduction}
Structured light \cite{forbes2021structured} i.e., tailoring various degrees of freedom light with a purpose has fueled numerous applications over the past decade \cite{he2022towards}, be it free-space or fiber communications \cite{al2021structured,wang2021generation}, high-dimensional quantum cryptography \cite{otte2020high}, optical trapping \cite{yang2021optical} or microscopy\cite{maurer2011spatial}. The linear manipulation of structured light has become ubiquitous of late with the advent of rewritable computer-controlled devices such as spatial light modulators \cite{yang2023review,rosales2024structured}, and digital micro-mirror devices \cite{scholes2020structured,perumal2023broadband}, while metasurfaces have offered unprecedented functionality \cite{dorrah2022tunable}. 

Recently, the focus has shifted towards nonlinear control of structured light \cite{buono2022nonlinear, li2017nonlinear}.  Initially demonstrated with second harmonic generation with orbital angular momentum (OAM) \cite{dholakia1996second} it has more recently been used to challenge our paradigms in nonlinear optics, e.g., away from just frequency conversion. In the context of structured light, nonlinear interactions have given rise to high-dimensional quantum teleportation without ancillary photons  \cite{sephton2023quantum,qiu2023remote}, real-time aberration correction without measurement \cite{singh2024light}, light diffracting off light \cite{da2022observation}, lock-and-key encryption \cite{xu2023orthogonal}, nonlinear holography  \cite{ackermann2023polarization, liu2020nonlinear}, novel imaging techniques \cite{qiu2018spiral,hong2020second,wang2021mid,sephton2019spatial}, vectorial nonlinear optics \cite{wu2019vectorial,wu2022conformal},  and even casting shadows with light \cite{abrahao2024shadow}. Nonlinear processes have been instrumental as a structured light creator and detector \cite{steinlechner2016frequency}, for instance, beam shaping in 2D \cite{shapira2012two} and 3D \cite{wei2019efficient},  and in modal detection schemes \cite{sephton2019spatial}. In all the aforementioned applications, the structure of the input and output must be carefully considered for the efficiency of the process \cite{singh2023frequency}, while spatial overlap and mode choice are understood to be crucial in determining the modal fidelity, with even small amplitude and phase perturbations introduced by the structure of the light itself having deleterious effects \cite{pinheiro2022spin,pereira2017orbital}.

Here we embrace the notion that light can diffract off light inside a nonlinear medium and introduce the idea of a virtual structured (in amplitude and phase) light-based aperture.  This allows us to use light to align light, where the structure of one mode is used to create our virtual light-based aperture. To demonstrate the concept we use difference frequency generation (DFG) and a set of amplitude (cross-hairs and sharp-edged dark modes) and phase (optical vortices) as our apertures to maximise the modal overlap of two OAM structured input light beams. This gives us control over the transverse cartesian and polar coordinates of the field, thus enabling alignment of the lateral, radial, and azimuth components. We show how both the transverse and longitudinal overlap can fine-tuned for final DFG fidelities exceeding $90\%$.  Our results can easily be generalised to arbitrary structured light patterns and other nonlinear processes, and will enhance emerging nonlinear applications where the structure of light plays an essential role.

\section{Concept and Theory} 
\begin{figure}[h!]
\centering\includegraphics[width = 12cm , keepaspectratio =true]{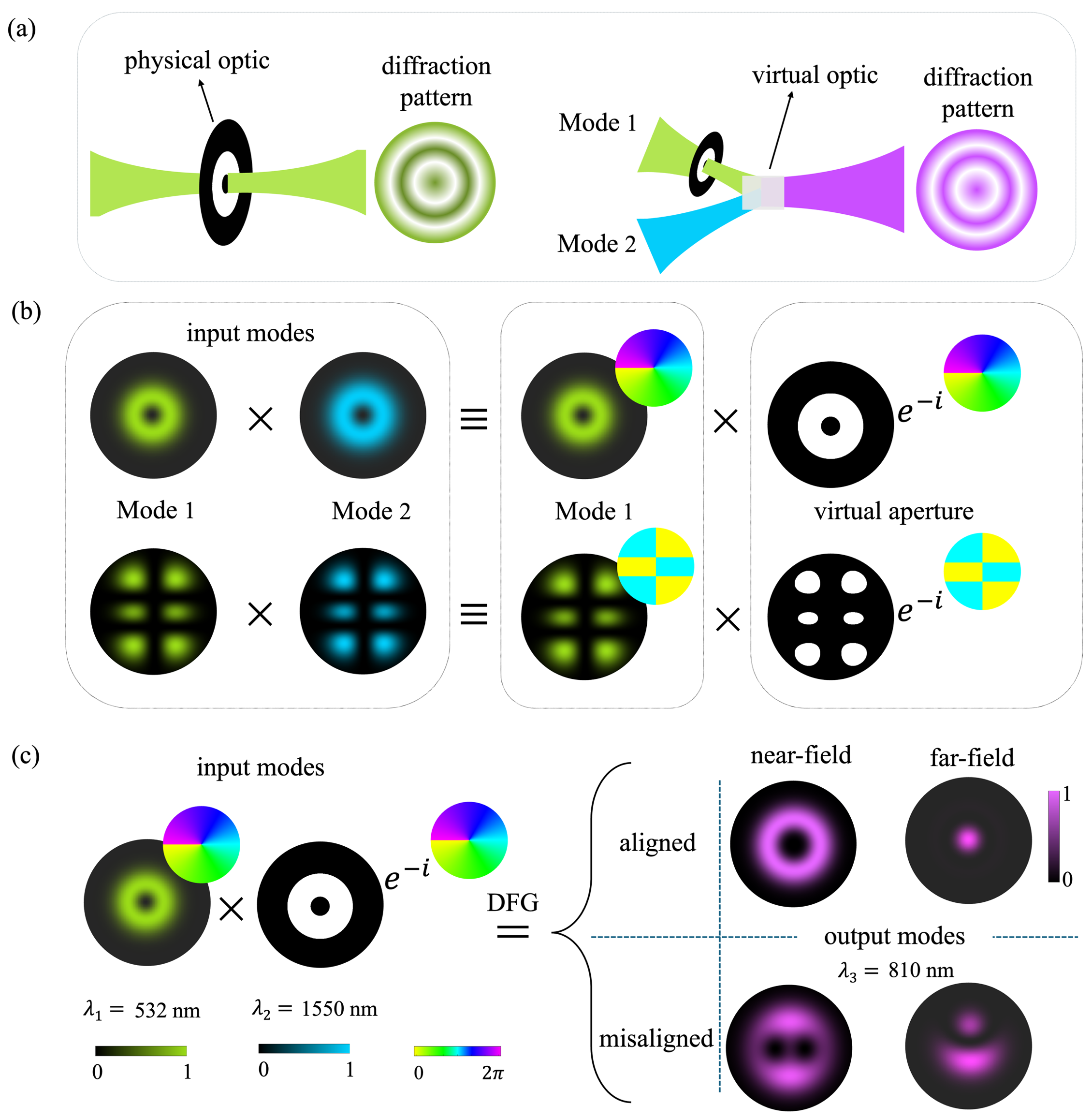}

 \caption{(a) The diffraction pattern from a light-based virtual optic is the same as that of a physical optic, but at a new wavelength. (b) Mode 2 acts as a transmission function for Mode 1 due to the product relationship in DFG, shown here with OAM and HG examples. (c) An $l = 1$ OAM input sees another as a virtual aperture. When aligned, the near-field (NF) is a ring-like structure and the far-field (FF) a Gaussian-like spot. When misaligned, the structure changes to reveal split vortices in the NF and a complex pattern in the FF.  
 }
\label{fig:concept_align}
\end{figure}

To illustrate the concept, we consider the scenario where two input modes are overlapped in a nonlinear crystal to produce a new structured light beam by DFG. The goal is to demonstrate that this can be leveraged to achieve an effect similar to that observed when light fields interact with physical apertures characterized by a transmission function, $t_\text{optic}$, as illustrated in Fig. \ref{fig:concept_align} (a).  
With consideration to the experiment to follow, we use the example where the DFG output is generated at a wavelength $\lambda_3 = 810$ nm from two input wavelength modes of $\lambda_1 = 532$ nm and $\lambda_2 = 1550$ nm.   However, to emphasise our approach, the wavelength conversion in the process will be largely ignored, and we focus our attention instead on the structure of the inputs and output, using the nonlinear process as a shaping tool \cite{steinlechner2016frequency}. Because the electric field of the DFG output is proportional to the product of the two inputs (with one mode conjugated), $E_\text{DFG} \propto E_1 E_2^*$,  as illustrated in Fig. \ref{fig:concept_align} (a), we have the notion of light diffracting off light, where one mode (Mode 1) can be viewed as the optical transmission function that the other mode (Mode 2) is modulated by, as in $E_\text{DFG} \propto E_1 t_\text{optic}^*$ or equivalently $E_\text{DFG} \propto E_2^* t_\text{optic}$.  

While the aperture in Fig. \ref{fig:concept_align} (a) is amplitude-only, in general it can be complex. We show this in Fig.~\ref{fig:concept_align} (b) for the example of two OAM modes of topological charge $l = 1$ as well as two Hermite-Gaussian modes, the latter to make clear that this procedure is not OAM specific nor restricted to modes of specific symmetries.  One of the modes can be expressed as a virtual optic for the other, with some amplitude and phase structure. We therefore introduce the notion that in a nonlinear process, light can be used to align light, using the structure of one mode as a virtual light-based aperture for the other.  This is important since any deviation from Gaussian modes has two major drawbacks: (i) structured apertures rather than simple pinholes are needed for filtering and aligning \cite{pinnell2020spatial}, and (ii) arbitrary structured light will not be the Fourier transform of itself, so the near-field (NF) and far-field (FF) patterns may be substantially different \cite{sroor2021modal}.  We will show that the versatility of our approach allows us to overcome both these challenges.

Consider the example of two $l = 1$ OAM modes as shown in Fig.~\ref{fig:concept_align} (c). The DFG pattern can now be understood as an interaction between one OAM beam of charge $l = 1$ and an annular spiral phase plate of charge $l = -1$. If the modes are perfectly aligned, then the NF is a ring of light and the FF is a Gaussian-like spot, both with no OAM.  However, if one mode is not perfectly overlapped with the other, then the diffraction of the light will result in more complex structured patterns in the two planes, dictated by diffraction and OAM interactions in nonlinear crystals, e.g., OAM adds in sum frequency generation but subtracts in DFG (see Ref.~\cite{wu2023observation} for an interesting deviation from the well-understood selection rules \cite{buono2022nonlinear}).  However, we are free to design the virtual light-based aperture at will by simply shaping one of the two input modes, a task easily achieved with standard structured light shaping tools.  This should be done judiciously based on the relevant structure of the modes and the desired outcome.  If the virtual aperture is known, then the undesired structure at the output can be used to infer information on the lateral and longitudinal overlap by simple diffraction considerations.  We will use some example apertures to reveal the benefit of the approach. 

\section{Experiment}
\begin{figure}[t]
\centering\includegraphics[width = 1\linewidth]{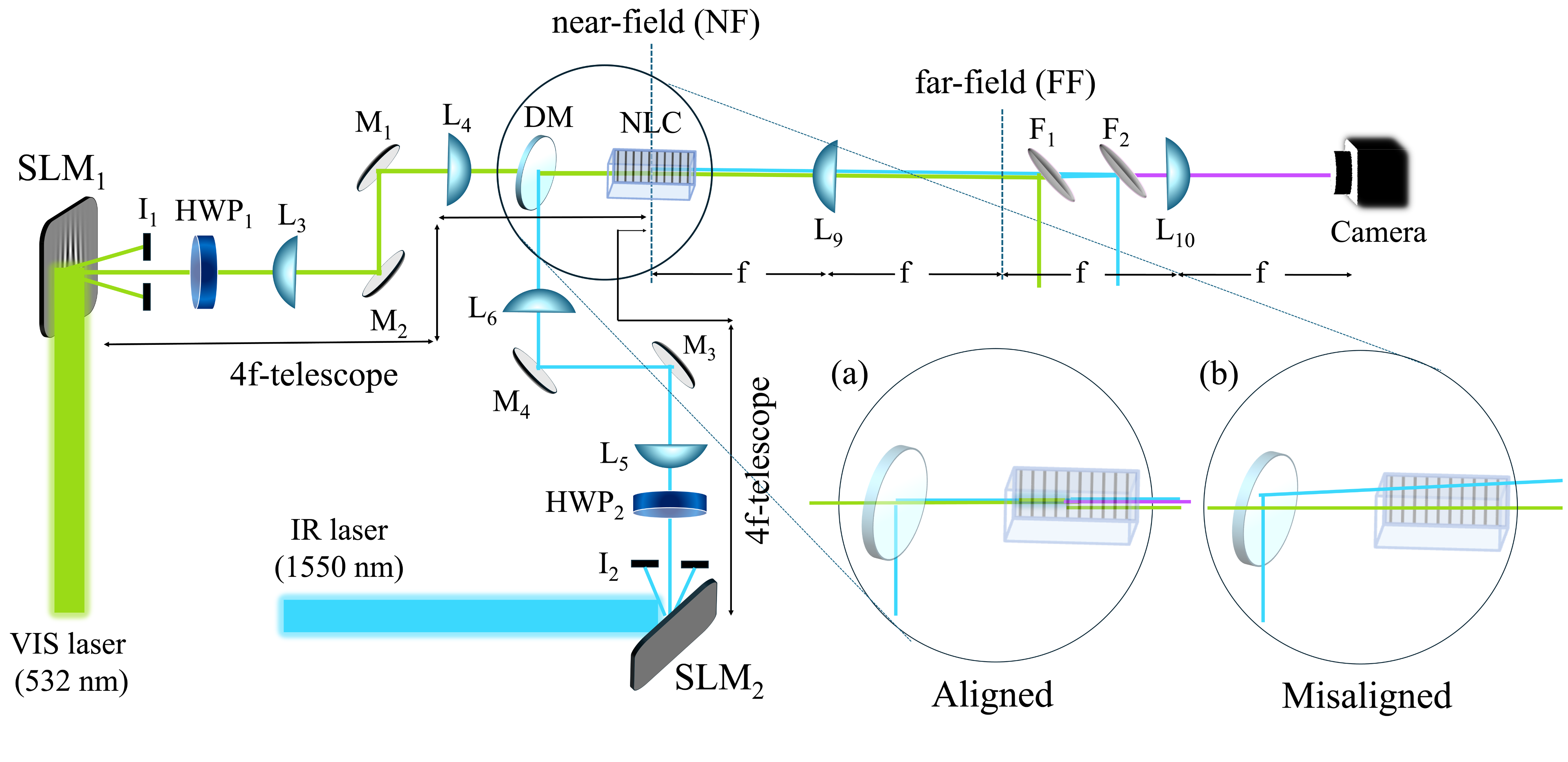}
\caption{The VIS input beam with wavelength 532~nm and IR beam 1550~nm is imaged using two 4f-telescopes at the nonlinear crystal (NLC) plane P. The output DFG from the crystal is imaged at a 4f distance from NLC to the detector plane or near-field plane, far-field  is measured using a camera put one focal length away from the lens L$_5$. Inset (a) illustrates the scenario where two input beams perfectly overlap at the crystal plane. (b) depicts the complete misaligned case with no spatial overlap. HWP: half-wave plate; I: aperture; DM: dichroic mirror; SLM: spatial light modulator; NF: near-field plane; FF: far-field plane; M: mirror; L: plano-convex lenses. \label{fig:setup}
}
\end{figure}

\begin{figure}[hb!]
\centering\includegraphics[width = 14cm , keepaspectratio =true]{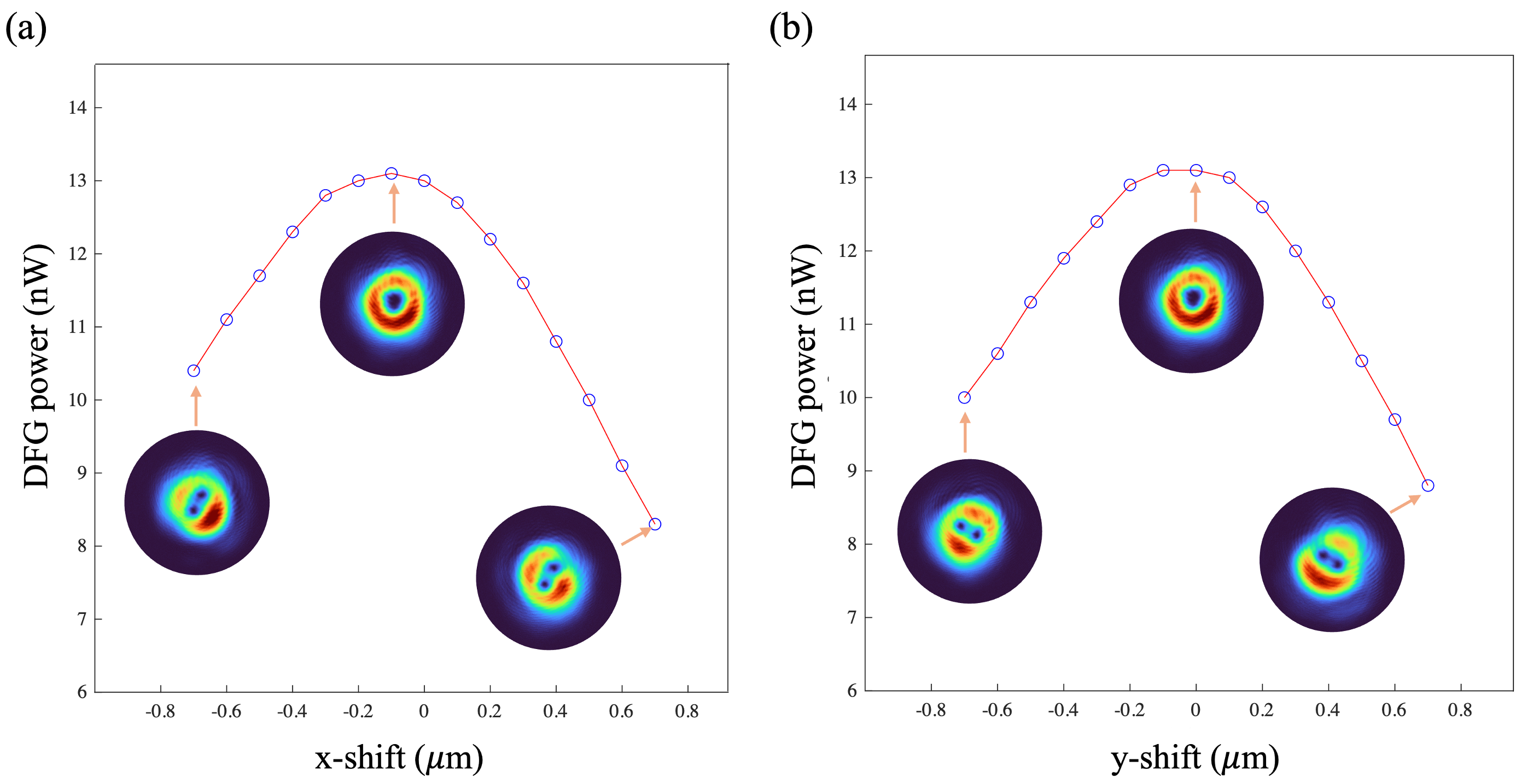}
\caption{(a) Output DFG power with change in the VIS beam position with hologram shift in x-direction (horizontal).
(b) The change in the output DFG power with a y-direction (vertical) shift in hologram position.  } \label{fig:concept_shift_vs_power}
\end{figure} \label{fig:efficiency}

The nonlinear structured light experiment is based on type-0 DFG with a 5~mm long periodically poled KTP crystal setup in a collinear configuration as shown in Fig.~\ref{fig:setup}. Two input beams with wavelengths $\lambda_1 = 532$~nm (VIS) and $\lambda_2 = 1550$~nm (IR)  were combined at the nonlinear crystal (NLC) under non-critical quasi-matching configuration (temperature tuned), generating an output DFG beam at $\lambda_3 = 810$~nm. The collimated VIS and IR input beams were modulated by spatial light modulators (SLM$_1$ and SLM$_2$) encoded using complex amplitude modulation\cite{carmelo2017}. The desired beam pattern in the first diffraction order was separated from undesired orders with the help of pinholes I$_1$ and I$_2$. The two beams were combined using a dichroic mirror (DM) and imaged at the NLC. To satisfy the product relationship between input modes (see appendix), it is crucial that the two beams should fall at the same position inside the NLC marked as NF. This was ensured by constructing two 4f-telescopes for each of the input beams. Mirrors M$_1$ and M$_2$ allowed precise control of the position and angle of the VIS beam within the crystal and similarly mirrors M$_3$ and M$_4$ on the IR side. The mechanical movement of mirrors was used to correct misalignment at the NLC. Fig.~\ref{fig:setup} (a) (inset) depicts the case of perfect alignment where two input beams completely overlap inside the crystal, resulting in a new frequency output beam. Fig.~\ref{fig:setup} (b) shows the extreme case of misalignment with no DFG, where two input beams are off both in angle and position despite passing through the crystal.  A pair of low-pass and high-pass filters were used to filter unconverted VIS and IR beams. The output DFG beam was imaged onto a camera setup at 4f distance away from the NF plane, allowing the observation of NF modes. For imaging FF modes, the camera was moved at a position one focal length away from the lens L$_9$. The DFG mode was further filtered by a bandpass filter ($\sim$10~nm) with a central wavelength of 810~nm mounted on the camera. The periodically poled crystal allows the two input beams to be of the same polarization (type-0 DFG). This was achieved by rotating the individual polarization directions using half-wave plates HWP$_1$ and HWP$_2$ and obtaining maximum DFG power from the crystal. The crystal temperature was set around 38$^\circ$C which was placed inside an electrically controlled oven having a stability of 0.01$^\circ$C.

\section{Transverse overlap}

Both the power of the output DFG mode and its structure are decided by the spatial overlap between the two input modes \cite{singh2023frequency}. Manual or digital coarse alignment is first implemented in the usual manner with tuning mirrors (or digital grating on the SLMs), taking into account practical issues such as temperature tuning of the crystal. A power measurement can be used to optimise the routine in a quantitative manner, as shown in Fig.~\ref{fig:efficiency}, if the modes share a common structure, but this is a special case and not the norm.  To demonstrate this, we created two OAM modes of the same intensity structure ($|l| = 1$) and shifted the hologram position of the visible pattern digitally on the SLM while measuring the DFG output power. As depicted in Fig.~\ref{fig:efficiency} (a), a small horizontal ($x$) shift of the order of a micrometre in the beam position at the NLC drops the output power by almost 50$\%$.  A similar trend is observed when the beams were moved in the vertical direction ($y$) as shown in Fig.~\ref{fig:efficiency} (b).

\begin{figure}[htbp]
    \centering\includegraphics[width = 1\linewidth]{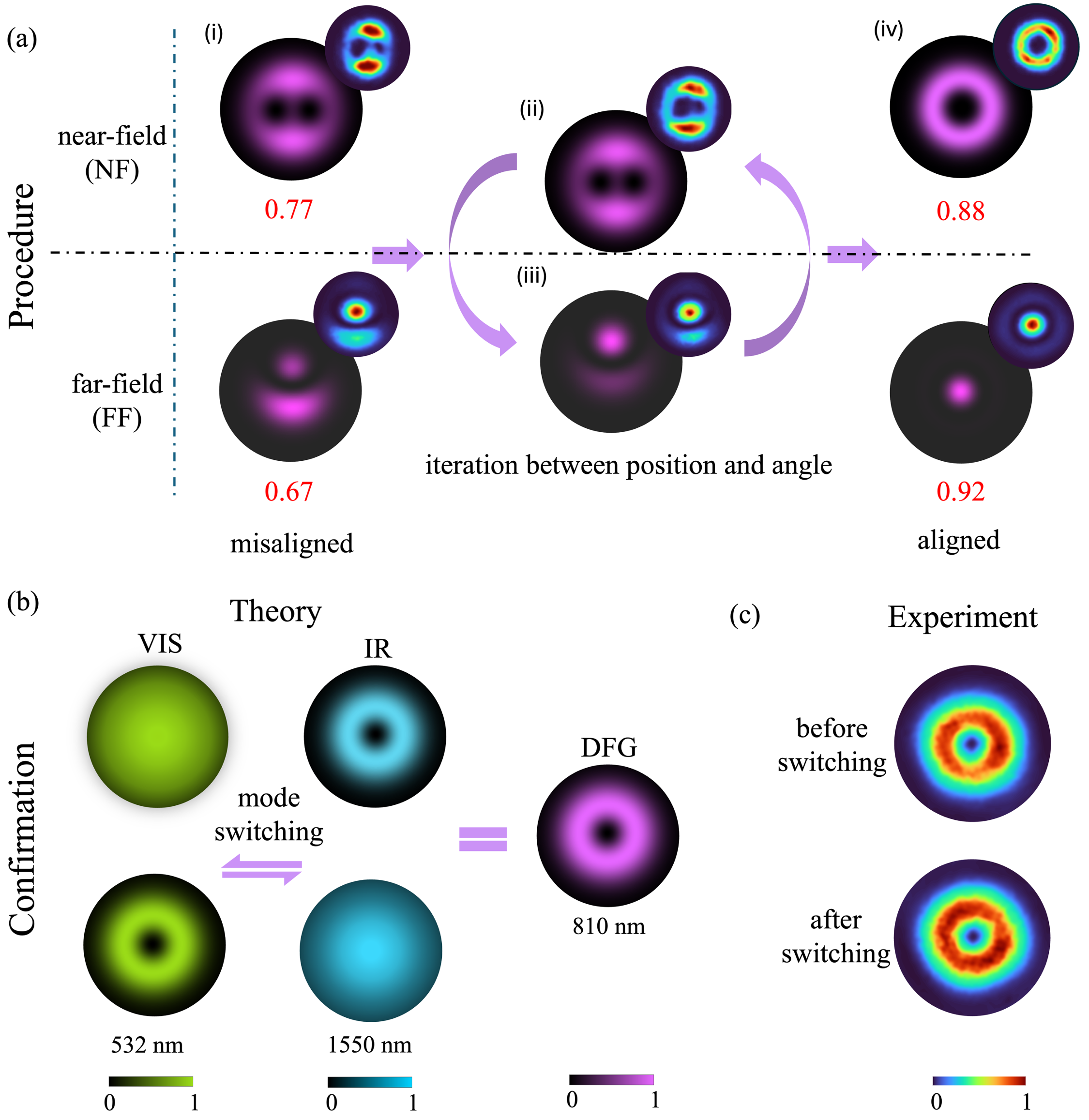}
\caption{(a) The alignment procedure with simulated and experimental results (insets) at both near-field (NF) and far-field (FF) planes. a(i) shows vortex splitting in the NF plane indicating the partial overlap (misalignment) at the nonlinear crystal (NLC) with a ringed Gaussian pattern in the FF plane. For misalignment correction, the positions and angles of the VIS (green) and IR (blue) beams are controlled using mechanical mirrors resulting in smaller and smaller singularity separation (a(ii))  and simultaneously enhancing the brightness of central Gaussian structure in the FF shown in a(iii). a(iv) presents the aligned case when two singularities merge completely resulting in a symmetric ring structure in the NF and a bright Gaussian structure with faint rings around it in the FF. The numbers below the panels represent the measured mode fidelity. (b) Confirmation of alignment by switching the Gaussian and OAM modes on both VIS and IR beams which should ideally result in the same DFG mode in the FF. (c) The experimental FF mode profiles before switching the modes and after switching the modes, a nearly identical structure confirms the good spatial overlap of two beams at the NLC.} \label{fig:iterative_alignment}
\end{figure}

But we note from the insets that the structure of the DFG also contains useful information.  Since the input modes are $l = 1$ and $l = -1$, the DFG output has an OAM of $l = 2$. In this instance, the $l = 2$ vortex splits into two $l  = 1$ vortices when the modes are imperfectly overlapped, a very sensitive measure of disturbance \cite{cheng2025metrology}.  This allows a new approach to improving the spatial overlap in the transverse plane based on the structure of light itself - iterating the alignment while observing modal structure rather than modal power, as shown in Fig.~\ref{fig:iterative_alignment} (a). Because the structure will differ from the NF to the FF, both planes hold useful information.  In this instance, split singularities in the NF plane and an asymmetrically structured bright-ringed Gaussian in the FF plane, as shown in top and bottom panels of Fig.~\ref{fig:iterative_alignment} (a), with experimental images of the outcomes shown in the insets.  Instead of power, we compare the theoretical and experimental light structures using fidelity (see appendix) as a correlation measure, with values ranging from 0 to 1, representing imperfect to perfect correlations, respectively. With coarse alignment (misaligned), the average fidelity is $\approx$ 0.7. When the separated singularities are made to merge in the NF plane the fidelity increases to nearly $0.9$ . The structure of the FF plane also contains useful information, as shown in Fig.~\ref{fig:iterative_alignment}, with lateral misalignment indicated by asymmetric structure.  By iterating the alignment based on both the NF and FF structure, the output beam fidelity increases to just over 0.9, a jump of almost 40$\%$.

For final confirmation of alignment, we generate a DFG beam with OAM ($l = 1$) mode in the FF plane. This can be done by encoding a large Gaussian beam on one wavelength and OAM mode on the other. The product rule suggests that the DFG profile should be independent of wavelengths, thus one can switch which input is Gaussian and which has OAM, i.e., which is the virtual light-based aperture. Ideally, both should result in an identical DFG pattern in the FF plane, shown in Fig.~\ref{fig:iterative_alignment} (b) and confirmed experimentally in Fig.~\ref{fig:iterative_alignment} (c). The circularly symmetric new frequency mode with smooth intensity distribution indicates a good transverse spatial overlap at the NLC.

\begin{figure}[h!]
    \centering
    \includegraphics[width = 14cm,keepaspectratio = true]{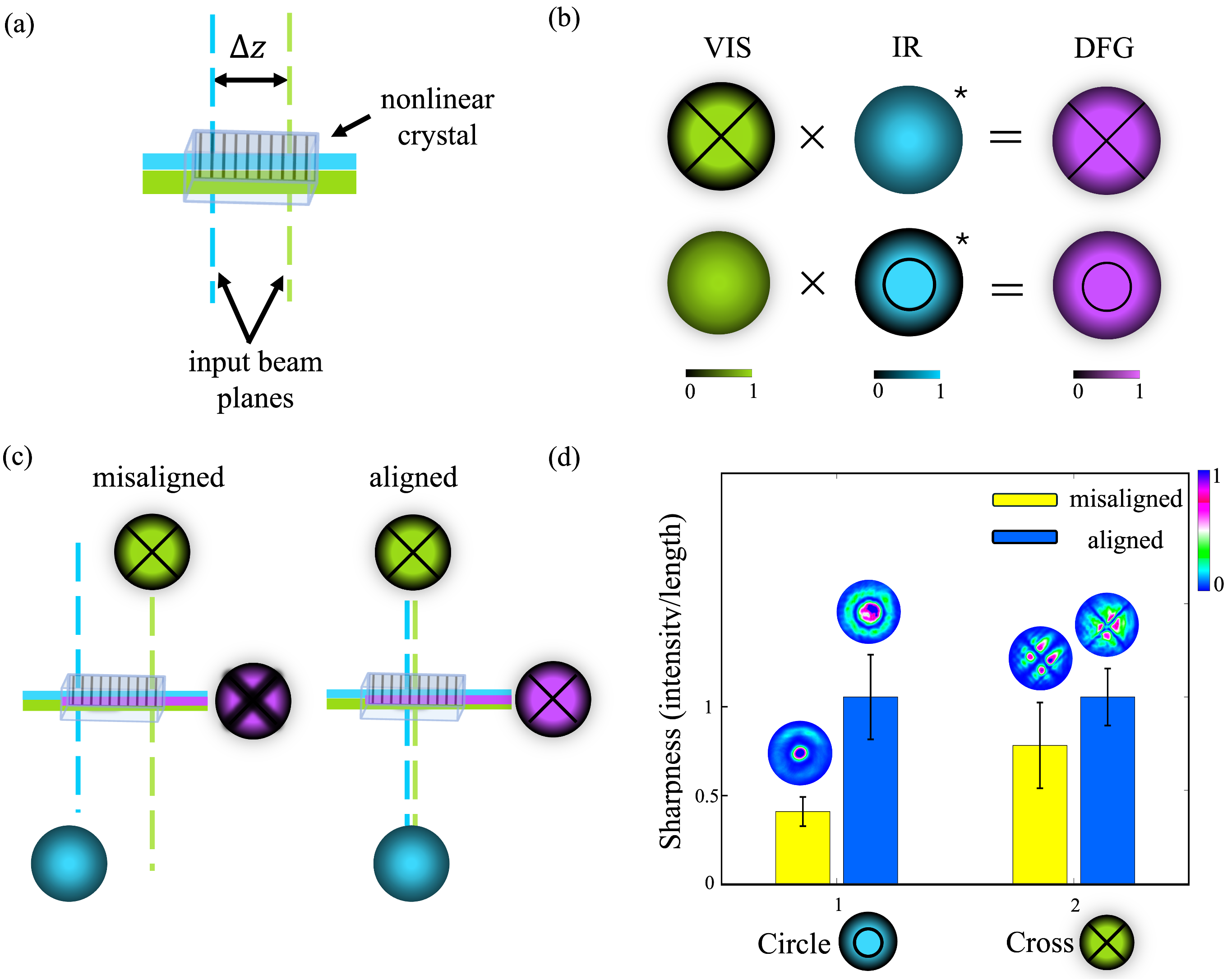}
    \caption{(a) The longitudinal overlap is reduced if the desired waist positions are displaced by some amount, $\Delta z$. (b) This is corrected by frequency conversion of sharp-edged modes with dark features, programmed sequentially on both modes to reduce wavelength-dependent effects. (c) The DFG sharpness is predicted to increase as the longitudinal overlap improves, shown for poor and good cases. (d) Experimental results, validating the procedure. 
    }
    \label{fig:z alignment} 
\end{figure}

\section{Longitudinal overlap}

As with lateral displacement, a longitudinal displacement, as illustrated in Fig.~\ref{fig:z alignment} (a), also reduces the modal overlap and modal fidelity of the output. We define the longitudinal displacement as the distance between the waist planes of the two input modes, $\Delta z$, but this can be recast to any user-defined definition based on the application at hand. To find and correct it, we used sharp-edged modes with dark features as our virtual apertures and a probe Gaussian beam, with the concept shown in Fig.~\ref{fig:z alignment} (b). We defined the sharpness as the change in intensity around the dark feature and alter on which mode the structure is encoded in order to reduce any wavelength-dependent effects, e.g., due to diffraction. As illustrated in Fig.~\ref{fig:z alignment} (c), when there is a displacement between the input planes, the aperture image is blurred, while as the displacement is reduced by adjusting the delivery optics or programming a lens function onto the SLMs, so the sharpness of the DFG structure increases.  Experimental results are shown in Fig.~\ref{fig:z alignment} (d), revealing a clear rise in sharpness due to improved longitudinal overlap. Notably, the technique is equally applicable for non-collinear configurations.

\begin{figure}[h!]
    \centering
    \includegraphics[width = 13cm,keepaspectratio = true]{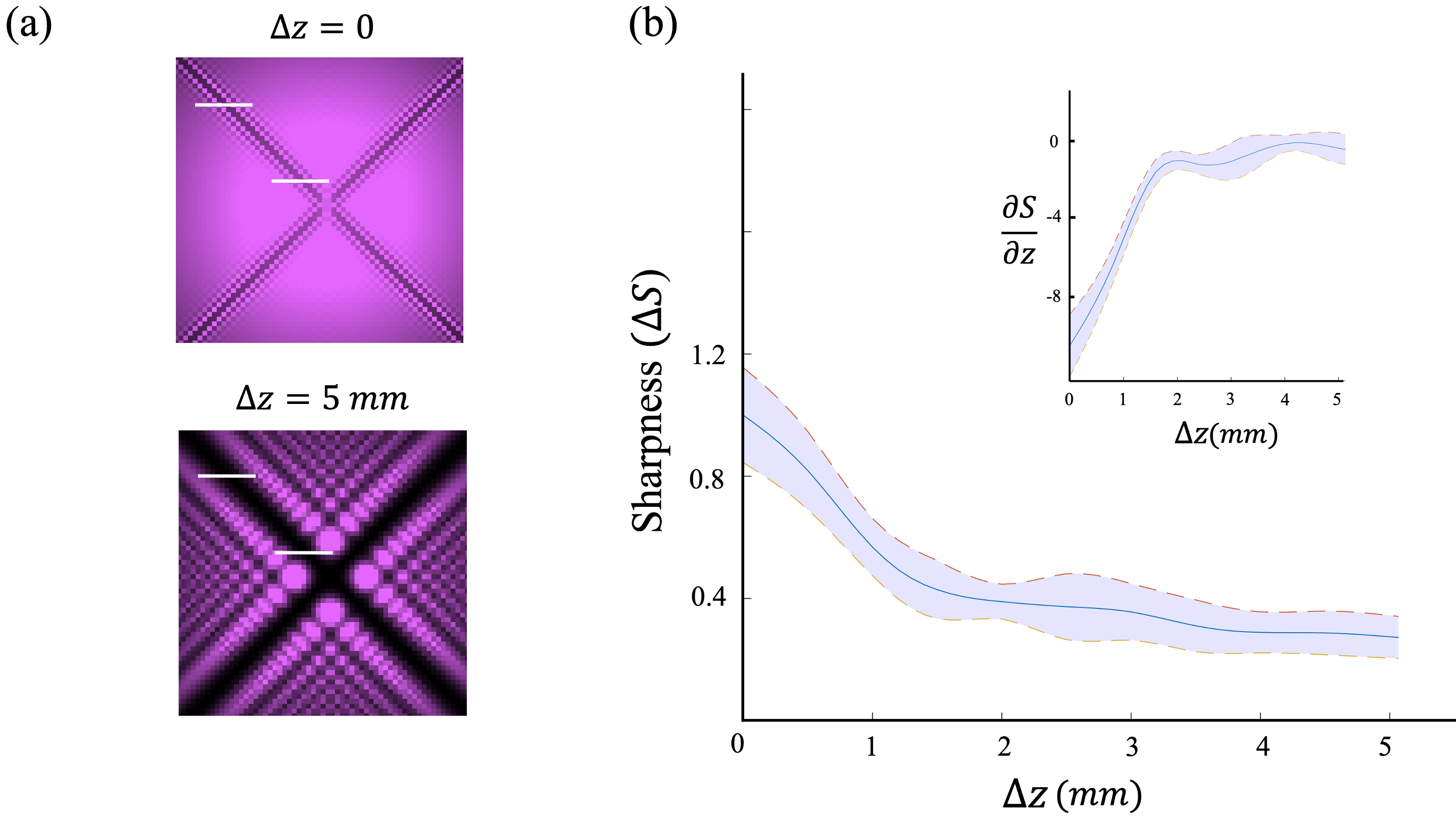}
    \caption{ (a) The simulated DFG profiles for longitudinal displacements $\Delta z = 0 $ and $\Delta z = $ 5~mm. (b) The decline in sharpness and inset plots the sensitivity of sharpness ($ \frac{\partial S}{\partial z} $) with the increase in $\Delta z$.}
     \label{fig:sharpness vs z} 

\end{figure}

It is instructive to consider how small a value of $\Delta z$ can be discerned by this approach. We answer this by numerical simulation, propagating the input beams away from their waist positions, with one longitudinally displaced by $\Delta z$ relative to the other. Since the sharpness depends on where the measurement is made on the resulting field, we took the average of 10 different spatial locations. Fig.~\ref{fig:sharpness vs z} (a) shows DFG structure for no displacement and 5~mm, with cross-sections around dark feature at two locations. To first approximation a change of sharpness, $\Delta S$, is related to the displacement by $\Delta S = \frac{\partial S}{\partial z} \Delta z$, shown in Fig.~\ref{fig:sharpness vs z} (b). The inset plots the factor $\frac{\partial S}{\partial z}$, which determines the sensitivity of sharpness to a change in displacement, to the displacement itself.  As expected, there is a general decline in sharpness as the longitudinal displacement increases.  More pertinently, the rate of change increases rapidly at small displacements.  This highlights that the approach is suitable for fine tuning and optimising the longitudinal overlap.  Similar trends were measured for the circular-edged aperture.

\section{Conclusion}
In conclusion, we demonstrated the use of light to align light for efficient and high-fidelity frequency conversion of structured light. To do so, we used the structure of one mode to create our virtual light-based aperture for the other, which we showed theoretically and experimentally for a range of structured light-based apertures acting on OAM modes as examples. This approach improved the lateral and longitudinal overlap of the modes, significantly enhancing the fidelity of the DFG output.  The technique relies on a product relationship between the two input modes at the crystal plane, which is an inherent feature of three-wave mixing processes, but could be extended to other nonlinear process too. The presented method will be a useful tool for high-fidelity frequency conversion of structured light in nonlinear optical systems, paving the way for advanced applications in biological imaging, spectroscopy, and optical communications.

\section*{Appendix}
The nonlinear polarization wave term for DFG  is proportional to the product of input waves i.e.,
\begin{equation*}
P({\omega_3}) =  2 \epsilon_o \chi^{(2)} E_1 \,E_2^{*}
\end{equation*}
This suggests a product relationship between the complex amplitudes of the input beams. For a paraxial nonlinear DFG system, under no pump depletion and optimum phase matching, the expression for waves  $E_2$ and $E_3$ is given as \cite{boyd_nonlinear},
\begin{align*}
\centering
    &E_2 = E_2(0)\, \text{cosh}\, \kappa z ,  \quad\quad  \kappa = \frac{\chi^{(2)} \omega_2 \omega_3  |E_1|}{\sqrt{k_2 k_3}c^2}     \\
    &E_3 = i \left( \frac{n_2 \omega_3}{n_3 \omega_2} \right)^{1/2}  \frac{E_1}{|E_1|}E_2^*(0) \text{sinh}\,\kappa z,    \label{Eoutput}   
\end{align*}
where $E_1, \, E_2$ and $E_3$ represent the complex amplitudes for 532~nm (VIS), 1550~nm (IR), and 810~nm (NIR) wavelengths respectively. We assume that the source planes of the two input beams fall at the center of the NLC, i.e. $z = 0$. To achieve it, we used two 4f-systems (Fig.~\ref{fig:setup}), for each of the input wavelengths, translating the source plane from SLM to the crystal interaction plane ($z = 0$). Thus, at the interaction plane under the limit $z \rightarrow 0$ and using approximation sinh\,$\kappa z \approx \kappa\, z  $ , the equations see the following, 
 \begin{gather*}
     E_3 \approx  i \left( \frac{n_2 \omega_3}{n_3 \omega_2} \right)^{1/2}  \frac{E_1}{|E_1|}E_2^*(0) \left(\frac{\chi^{(2)} \omega_2 \omega_3  |E_1|}{\sqrt{k_2 k_3}c^2}\right)  z . \\
 \end{gather*}
 and the complex amplitude $E_3$ at the crystal interaction plane can be simplified as,
 \begin{gather}
        E_3  \approx \eta E_1 E_2(0)^*, \quad \quad  \, \eta = i \frac{\chi^{(2)}  \omega_3}{ n_3 c} z  . \label{eq:product}
\end{gather}
At the output end of the crystal i.e. $z = L/2$, Eq. (\ref{eq:product}) turns into a similar expression as given in \cite{de2021beyond}. It is to be noted that the conjugate is on the beam with longer wavelength, which in our experiment is for 1550~nm. In case, the two input wavelengths carry different topological charges say $l_1$ and $l_2$, it can be easily seen that the OAM for DFG ($l_3$) should be the difference of the input OAMs, more specifically,
\begin{gather*}
    l_3 = l_1 - l_2
\end{gather*}
The output mode equal to the product of input modes (Eq. (\ref{eq:product})) is equally true for SFG under the limit $z \rightarrow 0$, where sinh\,$\kappa z $ is replaced by sin\,$\kappa z $. This implies that the product relationship between the input modes is valid close to $z = 0$ plane. The $z = 0$ can be an arbitrary plane at the NLC, meaning the source plane of all three beams should fall at the same position inside the NLC. This demands precise positioning of modes inside the NLC to correct for the spatial overlap in longitudinal direction (z-direction). We assumed the $z = 0$ plane at the center of the NLC. In order to avoid unwanted diffraction effects the Rayleigh ranges ($z _R$) for the input beams are kept much larger than the length (L) of the NLC making the whole crystal fall in the NF region. We define fidelity as intensity correlations between simulation and experiment, using the formula below
\begin{equation*}
\text{Fidelity} = \frac{\left[\sum_{x,y}{\sqrt{I_{sim}(x,y)I_{exp}(x,y)}}\right]^2}{\sum_{x,y}I_{sim}(x,y)\sum_{x,y}I_{exp}(x,y)}
\end{equation*}
$I_{\text{sim}}$ and $I_{\text{exp}}$ represent the simulation and experiment intensities respectively. The simulation results are obtained using Eq. \ref{eq:product}.

\begin{backmatter}
\bmsection{Funding}
Sachleen Singh thanks the Wits-CNRS for scholarship support.  The authors thank SA QuTI and the CSIR Rental Pool Programme for financial and equipment support, respectively.

\bmsection{Acknowledgments}
The authors would like to thank Dr Isaac Nape for the initial discussions.
\bmsection{Disclosures}
\noindent The authors declare no conflicts of interest.
\bmsection{Data availability} Data underlying the results presented in this paper may be obtained from the authors upon reasonable request.
\end{backmatter}
%%%%%%%bibliography%%%%%%%%%%
\bibliography{final}
\end{document}